\documentstyle[aps]{revtex}

\begin{document}

\title{
\begin{flushright}
\vspace{-1cm}
{\normalsize MC/TH 97/13}
\vspace{1cm}
\end{flushright}
Renormalisation and power counting in effective field theories\\
for nucleon-nucleon scattering}
\author{Keith G. Richardson, Michael C. Birse and Judith A. McGovern}
\address{Theoretical Physics Group, Department of Physics and Astronomy\\
University of Manchester, Manchester, M13 9PL, UK\\}
\maketitle
\bigskip
\begin{abstract}
The renormalisation of NN scattering in theories with zero-range interactions
is examined using a cut-off regularisation and taking the cut-off to infinity.
Inclusion of contact interactions that depend on energy as well as momentum
allows the effective range to be fitted to experiment with any desired sign or
magnitude. However, power counting breaks down: terms of different orders in
the potential can contribute to the scattering amplitude at the same order.
\end{abstract}
\pacs{}

There has been much recent interest in the role of contact interactions in NN
scattering\cite{afg,ksw,md1,md2,kap,lm,md3,md4,lep,vank}. Such interactions
arise naturally in effective field theories (EFT's) based on extensions of
chiral perturbation theory to describe the low-energy interactions among pions
and nucleons\cite{dgh,bkm}. In particular, they appear in the programme,
suggested by Weinberg\cite{wein2} and implemented by Ordonez {\it et
al.}\cite{orvk}, for constructing a chiral expansion of the NN potential. When
the full scattering amplitude is calculated from such a potential using a
Schr\"odinger or Lippmann-Schwinger equation, these zero-range pieces have to
be regularised and renormalised before predictions can be made for physical
quantities.

Various schemes for achieving this have been suggested. One approach is to
replace the contact interaction with one of finite range, at a scale
corresponding to the physics that has been integrated out\cite{orvk,md3,lep}
(see also\cite{nrqed}). Most of the recent debate, however, has been generated
by the results obtained when the interaction is taken to be truly zero-range
and this divergent potential is formally iterated to all orders. This requires
that it be regulated, either by introducing a cut-off which is taken to
infinity at the end of the calculation\cite{afg}, or by using dimensional
regularisation (DR)\cite{ksw,kap,lm}.

The treatments using DR lead to rather different results from the cut-off
ones, as has been discussed in some detail by the Maryland
group\cite{md1,md2,md3,md4} and Lepage\cite{lep}. In particular, the DR
approach can describe scattering with a positive effective range, though
convergence is only obtained for momenta lower than $1/\sqrt{ar_e}$, where $a$
is the scattering length and $r_e$ is the effective range. In contrast, cut-off
regularisation cannot give a positive effective range when the cut-off is
taken above some critical point\cite{md1,md2,md3,md4}.

In the present paper, we examine in more detail the renormalisation of NN
scattering in theories with zero-range interactions, using a cut-off
regularisation and taking the cut-off to infinity. We find that including
contact interactions that depend on energy as well as momentum allows the
effective range to be fitted to experiment with any desired sign or magnitude.
Unfortunately, however, we also find that the organisation of the calculation
in terms of powers of the momentum breaks down.

To keep the discussion as simple as possible, we consider a theory with
contact-interactions only, which would, for example, be applicable to NN
scattering at very low energies where pions have been integrated
out\cite{ksw,md3}. To second order in the momentum expansion, the
(unregulated) potential takes the form:
\begin{equation} \label{eq:pot1}
V(k',k)=C_0+C_2(k^2+k'^2)+\cdots.
\end{equation}
In the effective Lagrangian, these correspond to the contact terms 
$(\overline\Psi\Psi)^2$ and $(\overline\Psi\Psi)(\overline\Psi\nabla^2\Psi)
+{\rm H.c.}$. 

If the one-loop diagrams based on this interaction are evaluated, the
following integrals arise:
\begin{equation}
\label{eq:loop}
\int_{0}^{\infty}\frac{q^{2n+2}dq}{E-{q^2\over2\mu}+i\epsilon},
\qquad n\leq 2,  
\end{equation}
where $\mu$ is the reduced mass. When regulated with a cut-off which is 
taken to infinity, these contain divergent pieces multiplying
powers of the energy, $E^m$ where $m\leq n$. 

The natural way to cancel such divergences is to allow the coefficients in
Eq.~(\ref{eq:pot1}) to depend on the energy, for example $C_0\to
C_{00}+C_{01}E$. Such terms have not normally been considered in EFT
treatments of NN scattering, where energy dependence is usually eliminated
using the equation of motion.  It is clear however that if a potential is
going to be used in loop diagrams where the internal nucleons can be far off
shell, the two terms $C_{01}E$ and $C_2(k^2+k'^2)$ are not equivalent.
Consider, for example, a one-loop diagram with an insertion of one of these two
terms and another of $C_{00}$. In the case of $C_{01}E$ this generates 
only the $n=0$ divergent integral of Eq.~(\ref{eq:loop}) multiplied by $E$.
$C_2(k^2+k'^2)$ gives the same term multiplied by $k^2$ but also the $n=1$
integral. As the cut-off is taken to infinity the bare couplings will be
renormalised differently for the two vertices.  For any finite number of
loops, the renormalised result can be arranged to have the same form on-shell,
where $k^2=k'^2=2\mu E$, which tends to obscure the genuine difference between
the two terms, but as we shall see the same is no longer true for the sum of
all loop diagrams that is generated by the Lippmann-Schwinger equation.

Energy dependence of the potential should not be too surprising since it
arises whenever degrees of freedom are eliminated from a Schr\"odinger
equation\cite{fesh}. Indeed energy-dependent contact terms are naturally
required to renormalise, for example, contributions to the two-pion exchange
potential. In calculations of this potential\cite{orvk}, such energy
dependence is usually ignored on the grounds that, for a typical pion momentum
$k$, the nucleon energies in the energy denominators are of higher order in
$k$. However, while this power counting holds for the finite pieces of the
interaction, it fails for the divergent ones. Since the integrals have
power divergences, even a small energy can appear as the coefficient of a
divergence requiring a corresponding counterterm.

In treating the scattering non-perturbatively, we find it convenient to work
with the reactance matrix, $K$, rather than the scattering matrix, $T$. The
off-shell $K$-matrix for $s$-wave scattering satisfies the Lippmann-Schwinger
equation
\begin{equation} \label{eq:lse}
K(k',k;E)=V(k',k;E)
+\frac{\mu}{\pi^2}{\cal{P}}\int_{0}^{\infty}q^2dq\,\frac{V(k',q;E)K(q,k;E)}
{{p^2}-{q^2}},
\end{equation}
where we have introduced $p=\sqrt{2\mu E}$. Note that our definition of $K$ 
differs from the more standard one\cite{newton} by a factor of $-\pi$. This
equation is similar to that for the $T$-matrix except that the Green's
function satisfies standing-wave boundary conditions. This means that the usual
$i\epsilon$ prescription for the integral over $q$ is replaced by the principal 
value (denoted by ${\cal P}$). As a result, the $K$ matrix is real below
the threshold for meson production. Its on-shell value is related to the 
cotangent of the phase shift, and hence to the effective-range 
expansion by
\begin{equation} \label{eq:ere}
\frac{1}{K\left(p,p;\frac{p^2}{2\mu}\right)}=-\frac{\mu}{2\pi}p\cot\delta
=-\frac{\mu}{2\pi}\left(-\frac{1}{a}+\frac{1}{2}r_{e}p^2
+\cdots\right).
\end{equation}
The corresponding expression for the inverse of the on-shell $T$-matrix
differs from this by the term $i\mu p/2\pi$, which ensures that $T$ is unitary
if $K$ is Hermitian.

We regulate the zero-range potential of Eq.~(\ref{eq:pot1}) by introducing the
separable form,
\begin{equation}\label{eq:potreg}
V(k',k;E)=f(k/\Lambda)\left[C_0(E)+C_2(E)\,(k^2+k'^2)\right]f(k'/\Lambda),
\end{equation}
where the form factor $f(k/\Lambda)$ satisfies $f(0)=1$ and falls off rapidly
for $k/\Lambda>1$. We also have allowed for a possible energy-dependence of
the potential by letting the coefficients $C_{0,2}$ be functions of energy.
The resulting potential has a two-term separable form and so the corresponding
Lippmann-Schwinger equation can be solved using standard
techniques\cite{newton}. The off-shell $K$-matrix obtained in this way is
\begin{equation} \label{eq:koff}
K(k',k;E)=f(k/\Lambda)\frac{1+\frac{C_2}{C_0}(k^2+k'^2)
+\frac{C_2^2}{C_0}\left[I_2(E)-(k^2+k'^2)I_1(E)
+k^2k'^2I_0(E)\right]}{\frac{1}{C_0}-I_0(E)-2\frac{C_2}{C_0}I_1(E)
-\frac{C_2^2}{C_0}\left[I_2(E)I_0(E)-I_1(E)^2\right]}f(k'/\Lambda),
\end{equation}
where the integrals $I_n(E)$ are given by
\begin{equation}\label{eq:qint}
I_n(E)=\frac{\mu}{\pi^2}{\cal{P}}\int_{0}^{\infty}\frac{q^{2n+2}f^2
(q/\Lambda)}{{p^2}-{q^2}}dq,
\end{equation}
with $p^2=2\mu E$ again. Essentially the same expression has been obtained by
Phillips {\it et al.}\cite{md4}, who also considered regulating the theory by
cutting off the momentum integrals. If we had followed that approach and used
$f^2(q/\Lambda)$ as a cut-off on the integrals in the loops, the only
difference would be that the overall factor of $f(k/\Lambda)f(k'/\Lambda)$
would not be present in Eq.~(\ref{eq:koff}).

By expanding the integrals (\ref{eq:qint}) in powers of the energy (or $p^2$),
we can extract their divergent parts:
\begin{equation} \label{eq:expand}
\frac{I_n(E)}{2{\mu}}=-\sum_{m=0}^{n}A_m\Lambda^{2m+1}p^{2(n-m)}
+\frac{F(p/\Lambda)}{\Lambda}p^{2(n+1)},
\end{equation}
where the dimensionless integrals $A_m>0$ and $F(p/\Lambda)$ are finite as
$\Lambda\rightarrow\infty$. For large values of the cut-off $\Lambda$, or
small energies, the final term in this expansion can be neglected. (In fact
this term vanishes identically for the sharp momentum cut-off used
in\cite{md4}.) In that case, the scattering length can be written
\begin{equation}\label{eq:scattlen}
a=\frac{\mu}{2\pi}K(0,0;0)=\frac{1}{4\pi}\,
\frac{1-\frac{2{\mu}C_{2}^2}{C_0}A_2\Lambda^5}
{\frac{1}{2{\mu}C_0}+A_0\Lambda+2\frac{C_2}{C_0}A_1\Lambda^3+
\frac{2{\mu}C_2^2}{C_0}\Lambda^6\left(A_1^2-A_0A_2\right)}.
\end{equation}
If this expression is to remain finite as the cut-off is removed, we must
renormalise the potential by allowing the bare constants $C_{0,2}$ to depend
on $\Lambda$. In particular, we need them to vanish as $\Lambda\to\infty$ as
\begin{equation} \label{eq:scale}
C_0(0)\sim \frac{1}{\mu\Lambda},\qquad\qquad C_2(0)\sim\frac{1}{\mu\Lambda^3},
\end{equation}
so that no term in the numerator diverges, and none in the denominator
diverges faster than $\Lambda$. A finite scattering length can then be obtained
if the leading terms of $C_0$ and $C_2$ are chosen so that the coefficient of
$\Lambda$ in the denominator is zero. This scale dependence of $C_0$ is that
same as that found by Weinberg\cite{wein2} and Adhikari and
coworkers\cite{afg} when they renormalised the scattering amplitude obtained
from the lowest-order potential, $V(k',k)=C_0$.

Since the leading terms in the denominator must cancel, the finite remainder
will depend on the sub-leading terms. This is most cleanly illustrated if we
assume that the coefficients $C_{0,2}$ depend analytically on $\Lambda^{-1}$ as
$\Lambda\rightarrow\infty$. In this case, it is convenient to introduce 
dimensionless coefficients $\alpha_i$ and $\beta_i$ by writing
\begin{equation}\label{eq:subl}
2\mu\Lambda C_0=\alpha_0+{2\mu\beta_0\over\Lambda}\qquad\hbox{and}\qquad
2\mu\Lambda^3 C_2=\alpha_2+{2\mu\beta_2\over\Lambda}.
\end{equation}
After taking the cut-off to infinity, we obtain 
\begin{equation}\label{eq:result}
{1\over K(k',k;E)}=4\mu^2{A_0\beta_0+\bigl(2A_1+2\alpha_2(A_1^2-A_0A_2)\bigr)
\beta_2\over \alpha_0-\alpha_2^2A_2},
\end{equation}
with the condition
\begin{equation}
1+\alpha_0A_0+2\alpha_2A_1+\alpha_2^2(A_1^2-A_0A_2)=0.
\end{equation}

Although this leads to a finite scattering length, we see that all explicit
energy and momentum dependence in the $K$-matrix (\ref{eq:koff}) vanishes.
Hence, under the assumption of analytic dependence on the cut-off, a non-zero
effective range can only be obtained as the cut-off is removed if the
coefficients in the potential (\ref{eq:potreg}) are allowed to depend on
energy. Either or both of the subleading coefficients $\beta_i$ may be given a
linear energy dependence to generate a finite scattering length and effective
range. This corresponds to a potential of the form
\begin{equation}
V(k',k;E)=f(k/\Lambda)\left[C_{00}+C_{01}E+(C_{20}+C_{21}E)\,(k^2+k'^2)\right]
f(k'/\Lambda).
\end{equation}
We observe from Eq.~(\ref{eq:subl}) that the cut-off dependences of $C_{01}$
and $C_{20}$ are different, indicating that these terms cannot be equivalent in
the limit $\Lambda\rightarrow\infty$.

The above analysis assumes an analytic dependence of the coefficients $C_0$
and $C_2$ on $\Lambda^{-1}$.  An alternative approach, suggested by Phillips
{\it et al.}\cite{md4}, involves instead obtaining simultaneous equations for
(energy-independent) $C_0$ and $C_2$ in terms of $a$ and $r_e$ from the
expansion to $O(p^2)$ of $1/K(p,p;p^2/2\mu)$ from Eq.~(6). At least for
$r_e\leq 0$, these equations can be solved for any $\Lambda$, yielding
expressions for the $C$'s in terms of the observables $a$ and $r_e$. These
expressions can be expanded in powers of $\Lambda^{-1/2}$, with the leading
behaviour given again by Eq.~(10). All terms up to order $\Lambda^{-2}$ beyond
the leading order must be kept to obtain a finite $a$ and a finite (but
negative) $r_e$ as $\Lambda\to\infty$. If however the matching is imposed only
as $\Lambda\to \infty$ we can add an expression of the form
$\Lambda^{-4}g(\Lambda^{-1})$ to $C_2$, where $g$ is some analytic function,
at the same time altering $C_0$ to ensure that $C_0-2\mu\Lambda^5A_2C_2^2 $ is
unchanged, without changing the scattering length or effective range.  (It
should be noted that though these extra terms vanish as $\Lambda\to\infty$,
they do so less fast than terms that do contribute.) Thus the relationship
between bare and renormalised parameters is not unique. Furthermore, energy
dependence of the $C$'s is still required to obtain a positive effective range.

Quite different results are obtained using DR. Since all the terms in the
expansions of the integrals $I_n(E)$ either diverge like a power of $\Lambda$
or vanish as $\Lambda\rightarrow\infty$, DR, being sensitive only to
logarithmic divergences, sets these integrals to zero. The resulting
expression for the on-shell $K$-matrix is thus given by the first Born
approximation\cite{ksw}. This involves only on-shell matrix elements of the
potential and so energy and momentum dependence cannot be distinguished.
Moreover, a large scattering length leads to a $K$-matrix that varies much
more rapidly with energy than the corresponding effective range expansion.
Hence DR leads to a potential with a very limited range of
validity\cite{ksw,lm,lep}, unless rapid energy dependence is generated through
the introduction of additional low-energy degrees of freedom\cite{kap}.

We have seen that, if one works with a cut-off regularisation and takes the
cut-off to infinity, energy-dependent terms in the potential both arise
naturally and have quite different effects from momentum dependence. In
particular they allow for a finite effective range of either sign. Although
they might seem to be the natural answer to the apparent difficulties of
contact interactions, there is however a drawback. In order to justify
truncation of the effective potential, some form of power counting has to be
valid.  Usually it is assumed that, provided the nucleon momentum $k$ is
small, the terms omitted from the potential will contribute at higher order in
$k/M$, where $M$ is some scale associated with the physics that has been
integrated out. In Eq.~(\ref{eq:result}), however, the two terms in our
potential both contribute to the scattering length, even though one is of
order unity and the other is of order $k^2$. Similarly, the effective range
receives contributions from terms in the potential of order $E$ and $Ek^2$.
Power counting has broken down, and there is no guarantee that the terms we
have omitted will not change our results. Given this lack of a consistent
expansion as the cut-off is removed, it is perhaps unsurprising that
renormalisation schemes based on cut-offs and DR should yield different
results.

The only alternative is not to take the cut-off $\Lambda$ to infinity, but to
set it to some scale corresponding to the physics that has been omitted from
the effective theory\cite{orvk,md3,lep}. This form of regularisation is
appealing, since we know that the physics described by the contact
interactions is not truly zero-range. In this approach, terms in the potential
of different orders can contribute to some scattering observable at the same
order, as can be seen from our expressions above. Nonetheless provided that
the coupling constants are natural, that is of order unity when expressed in
terms of $\Lambda$, this leads to an expansion of observables in powers of
$p/\Lambda$. Although for any given system one does not know {\it a priori}
whether such an expansion will be valid, the results are presumed to be
trustworthy so long as one can find a region where the predictions are not
overly sensitive to the cut-off\cite{lep}. Moreover if the cut-off is left
finite, then it may be possible to transform the fields and convert energy
dependence of the potential into momentum dependence (nonlocality), or {\it
vice versa} as has long been done in nuclear physics\cite{perey}. Hence, in
practical applications of this approach, energy-dependent terms of the sort
discussed here have not generally been considered\cite{orvk,md3,lep}.

\vspace{10pt}
{\bf Acknowledgements:} {We are grateful to Daniel Phillips and Bira van Kolck
for stimulating discussions. This work was supported by the EPSRC and PPARC.}

\end{document}